\documentclass{emulateapj}
\newcommand{\Mpc}{\mbox{ Mpc}}

\newcommand{\hunits}{\mbox{ km s$^{-1}$ Mpc$^{-1}$}}

\newcommand{\lya}{Ly$\alpha$ }

\newcommand{\bq}{\begin{equation}}
\newcommand{\eq}{\end{equation}}
\newcommand{\bqa}{\begin{eqnarray}}
\newcommand{\eqa}{\end{eqnarray}}
\def\VEV#1{\left\langle #1\right\rangle} 

\begin{document}

\title{Large-Scale Fluctuations in the \ion{He}{2} \lya Forest and \ion{He}{2} Reionization}

\author{Steven R.  Furlanetto \& Keri L. Dixon}

\affil{Department of Physics \& Astronomy, University of California Los Angeles; Los Angeles, CA 90095, USA; sfurlane@astro.ucla.edu}

\begin{abstract}
We examine large-scale fluctuations in the \ion{He}{2} \lya forest transmission during and after \ion{He}{2} reionization.  We use a simple Monte Carlo model to distribute quasars throughout a large volume and compute the resulting radiation field along one-dimensional skewers.  In agreement with previous studies, we find that the rarity of these sources induces order unity fluctuations in the mean optical depth after reionization, even when averaged over large segments ($\sim 10$--$100 \Mpc$ across).  We compare our models to existing data along five \ion{He}{2} \lya forest lines of sight spanning $z \sim 2$--$3.2$.  The large cosmic variance contained in our model plausibly explains many of the observed fluctuations at $z \la 2.5$.  But our models cannot accommodate the large fluctuations toward high optical depths on $\sim 30 \Mpc$ scales observed at $z \sim 2.7$--$2.9$, and the measured optical depths ($\tau_{\rm eff} \ga 4$) at $z>2.9$ are difficult to explain with a smoothly-evolving mean radiation field.  In order to better understand this data, we construct a toy model of \ion{He}{2} reionization, in which we assume that regions with the smallest radiation fields in a post-reionization Universe (or farthest from strong ionizing sources) are completely dark during reionization.  The observed fluctuations fit much more comfortably into this model, and we therefore argue that, according to present data, \ion{He}{2} reionization does not complete until $z \la 2.9$.
\end{abstract}
  

\section{Introduction} \label{intro}

Our premier tool for studying the intergalactic medium (IGM) is the \lya forest, from which we can measure the absorption properties of the low-density ``cosmic web" lying between galaxies.  Although observations of the \ion{H}{1} forest have a long and rich forty-year history \citep{rauch98}, recently there has been great progress in measuring absorption from other ions, including both heavy elements and helium, the second most common element in the Universe.

Unfortunately, observing the \ion{He}{2} \lya forest is difficult, because the transition's rest wavelength of 304 \AA \ places much of the absorption in the far ultraviolet, even for material at $z \sim 2$--$3$.  Not only do ultraviolet telescopes present formidable technical challenges, but studying the forest also requires a luminous far-UV source to provide a background light.  Such lines of sight are rare, because any that intersect strong \ion{H}{1} absorbers do not retain their UV flux.  To date, five quasars at $z \sim 2$--$3.5$ have satisfied these criteria and have high-resolution \ion{He}{2} \lya forest measurements (see \S \ref{data} below for references to the individual quasars), although very recently more than twenty other potential targets have been identified \citep{syphers09, syphers09b}.

Nevertheless, the known lines of sight are particularly interesting, because current data (from these measurements and other indirect techniques) suggest that quasars fully ionized \ion{He}{2} by $z \sim 3$.  Most obviously, there is a substantial decrease in the mean \lya forest absorption beyond $z \sim 2.9$ \citep{dixon09}, which is compatible with models where large swathes of \ion{He}{2} are cleared out by quasar light at about that time \citep{sokasian02, furl08-helium, mcquinn09}.  Indirect evidence from the IGM temperature also suggests reionization at $z \sim 3.2$--$3.4$ \citep{schaye00, ricotti00, lidz09-temp}, which is consistent with models of helium reionization ending at these slightly higher redshifts \citep{gleser05, furl08-igmtemp, mcquinn09}. 

Here, we will study the variability in the \ion{He}{2} \lya forest data and its consequences for reionization.  Specifically, recent work has shown that the high-energy radiation background fluctuates strongly both during and after helium reionization, thanks largely to the rarity of the bright quasars providing the ionizing photons (see \citealt{fardal98, maselli05, bolton06, meiksin07, tittley07, furl09-hefluc} for theoretical work and \citealt{shull04, fechner06, fechner07} for observations).  Moreover, these fluctuations can span large physical scales \citep{furl09-heps}.  We might therefore expect that cosmic variance from the small number of well-studied lines of sight may limit any firm conclusion based on the optical depth.  

On the other hand, the existing observations already exhibit substantial fluctuations and may themselves suggest something useful about reionization.  We naturally expect larger fluctuations during reionization, because the radiation field can be extremely large near quasars but almost zero in the \ion{He}{2} regions far from these sources \citep{meiksin07, furl09-hefluc}.  Here, we will quantify how these fluctuations impact measurements of the mean \lya forest absorption and show that the observed variations are by themselves powerful probes of the radiation field.  We will use a simple Monte Carlo model and ask whether the existing data are consistent with \ion{He}{2} reionization ending at $z \sim 3$.  Our approach is similar to \citet{fardal98}, although we use updated input parameters, add a toy model for reionization itself, and have the advantage of comparing to significantly more data.

This paper is organized as follows.  In \S \ref{mc}, we describe the Monte Carlo model that we use for the post-reionization limit.  We present a simple toy model of \ion{He}{2} reionization in \S \ref{reion} and illustrate the basic characteristics of these models in \S \ref{pdfs}.  In \S \ref{data}, we compare our results to existing data.  Finally, we conclude in \S \ref{disc}.

In our numerical calculations, we assume a cosmology with $\Omega_m=0.26$, $\Omega_\Lambda=0.74$, $\Omega_b=0.044$, $H=100 h \hunits$ (with $h=0.74$), $n=0.95$, and $\sigma_8=0.8$, consistent with the most recent measurements \citep{dunkley09}.  Unless otherwise specified, we use comoving units for all distances.

\section{Monte Carlo Model}
\label{mc}

\subsection{The Post-Reionization Limit}
\label{post-reion}

We have previously examined the distribution of ionizing background intensities in the post-reionization limit with an analytic model (\citealt{furl09-hefluc}; see also \citealt{zuo92a, meiksin03, meiksin07}).  This model predicts substantial line-of-sight variations in the specific intensity $J$, even on large scales, which can be quantified by, for example, the power spectrum of the ionizing intensity \citep{zuo92b, furl09-heps}.  However, the power spectrum does not perfectly match the method used to analyze the observations, where optical depth measurements are binned over large scales in order to reduce the (substantial) errors in measurements of individual features. These differences include: (1) real quasar absorption spectra allow only a finite dynamic range to measure transmission, (2) observations are typically binned with sharp filters in real space, (3) the (measurable) ionization rate requires an integral over the entire quasar spectrum, rather than evaluation at a single frequency, and finally (4) observations are sensitive to the density field and temperature of the IGM in addition to the radiation background studied by \citet{furl09-heps}.

To address (most of) these concerns, we instead use a simple Monte Carlo model to construct the ionizing field along example lines of sight.  Our model is very similar to \citet{fardal98}, except that we use more recent data as inputs to the calculation.  We begin by populating a box of size $\sim 1$~Gpc$^3$ with quasars by randomly sampling the \citet{hopkins07} luminosity function at a given redshift; the precise size of the box varies with the application but is always at least this large.\footnote{To generate all of our random numbers, we use the Mersenne Twister algorithm, which has a period of at least $2^{19937}-1$ \citep{matsumoto98}.}  We also draw the far-UV spectral index of each quasar (assumed to be a power law at $\lambda < 1050$~\AA) from a Gaussian distribution with $\VEV{\alpha}=1.5$ and variance $\sigma^2=1$, except truncated between $\alpha=(0.5,\,3.5)$.  This is roughly consistent with the observed distribution of \citet{telfer02}, and the central value is also consistent with the recent direct measurements of \citet{syphers09b}.  We assign quasar positions randomly throughout the box.  Note that this procedure does \emph{not} account for quasar clustering, which is substantial \citep{shen07} and will tend to amplify the fluctuations (see discussion in \S \ref{disc}). 

Next, we choose a line of sight of a specified length. To do this, we randomly choose the segment's starting position within the box and its direction on the unit sphere, wrapping the line of sight around the box if necessary.  We then compute the ionizing intensity in 1 Mpc steps along the line of sight.  In each case, we sum over all of the quasars in the box,\footnote{In detail, we use the closest position of each quasar in a periodic tiling of the box, so that we effectively include all sources within 1 Gpc of every point.  This scale is much larger than the assumed attenuation length, so more distant sources can safely be ignored.} so that the specific intensity at a given frequency (in units of erg cm$^{-2}$ s$^{-1}$ Hz$^{-1}$) is 
\bq
{J_\nu \over J_\star} = \sum_i {L_i \over L_\star} \left( {r_i \over R_0} \right)^2 e^{-r_i/R_0},
\label{eq:jsum}
\eq
where $J_\star = L_\star/(4 \pi R_0^2)$, $L_i$ is the quasar's luminosity, $L_\star$ is the mean luminosity, $r_i$ is the distance between the $i^{\rm th}$ quasar and the point of interest, and $R_0$ is the attenuation length of the ionizing photons (see below).

Finally, we are generally interested in the ionization rate $\Gamma$ rather than the specific intensity at a single frequency,
\bq
\Gamma = \int d\nu \, J_\nu \sigma_{\rm HeII}(\nu),
\eq
where $\sigma_{\rm HeII}$ is the photoionization cross-section.  If we approximate $\sigma_{\rm HeII} \propto \nu^{-3}$ relatively close to the ionization threshold \citep{verner96} and consider a quasar with spectral index $\alpha_i$, we have
\bq
{ \Gamma_i \over \bar{\Gamma} } = \left( { \VEV{\alpha} + 3 \over \alpha_i + 3} \right) \left( {4.6^{-\alpha_i} \over 0.0291} \right),
\eq
where $\Gamma_i$ is the ionization rate from the $i$th quasar, the last factor accounts for the spectral decline between 1050 \AA \ and the \ion{He}{2} ionization edge, and $\bar{\Gamma}$ denotes an average over the quasar distribution.  Note that here we have assumed that $R_0$ is \emph{independent} of frequency.  In reality, the attenuation length can depend strongly on frequency:  for example, $R_0 \propto 1/\sigma \propto \nu^3$ in a uniform medium.  The clumpiness of the IGM moderates this somewhat; for example, a column density distribution $\propto N_{\rm HeII}^{-1.5}$ implies $R_0 \propto \nu^{1.5}$ \citep{paresce80}.  Our simple treatment is only valid in the limit that all the absorption comes from extremely high-column density systems.  However, a more realistic approach will only \emph{decrease} the fluctuations and hence strengthen our conclusions, because it would increase the attenuation length at large frequencies and make the observed fluctuations even more difficult to explain.

For our calculations here, we generate three independent boxes at each relevant redshift. Within each box, we generate 100 lines of sight, each $\sim 1$~Gpc in length.  This procedure provides 3,000 samples of $\sim 100$ Mpc segments (or many more for comparisons to shorter segments).  We choose the precise dimensions to match the data points in the observed sample.  In practice, there is relatively little difference in the results between the three independent boxes.

\begin{figure}
\plotone{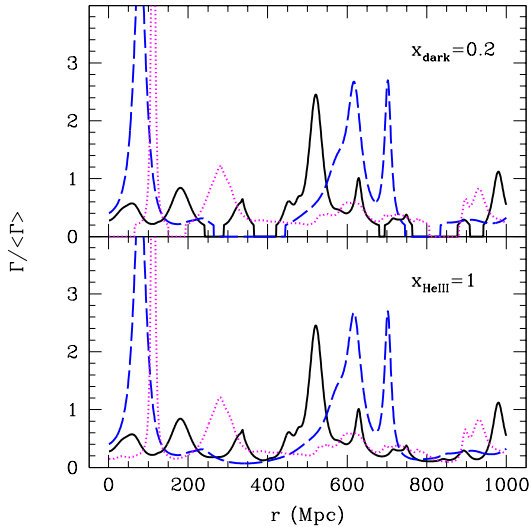}
\caption{Ionization rate (scaled to its mean) along three example lines of sight from our Monte Carlo simulations.  The bottom panel shows the post reionization limit; the top panel shows the same lines of sight processed with our toy model of \ion{He}{2} reionization, with $20\%$ of the pixels assumed to be outside of illuminated \ion{He}{3} regions.  All take $R_0=30 \Mpc$.}
\label{fig:examples}
\end{figure}

The bottom panel of Figure~\ref{fig:examples} shows three example lines of sight generated with this method; here, we have assumed $z=3$ and $R_0=30 \Mpc$ (see below).  We show the ionization rate scaled to its mean value, $\VEV{\Gamma} = 3 \bar{N}_0 \Gamma_\star$, where $\bar{N}_0$ is the mean number of quasars within one attenuation volume and $\Gamma_\star$ is the ionization rate for a point with specific intensity $J_\star$ \citep{meiksin03}.  These examples illustrate several important aspects of the signal (see also \citealt{furl09-heps}):  (1) the correlations can extend over extremely large scales, as in the long trough at $\sim 200$--$500 \Mpc$ in the dashed curve; (2) rare small-scale ``fluctuations" are present, because of the $1/r^2$ profiles around individual sources (these dominate the power spectrum); (3) the median of $J$ is a few times smaller than the mean; and (4) fluctuations far above the median level are relatively common, but the ionization rate never falls significantly below that level.  This last point is because the accumulated contributions of the many quasars at distances larger than $R_0$ provides an effective floor to $\Gamma$, and it is crucial for the following.

\subsection{From the Ionizing Background to the Effective Optical Depth} 
\label{tauflucs}

We now wish to transform our results to the observable \ion{He}{2} \lya optical depth.  Because we are interested in spatial fluctuations over relatively large scales, we will average the observed transmission ${\mathcal T}$ over large segments, parameterizing it in terms of the effective optical depth $\tau_{\rm eff}$,
\bq
\tau_{\rm eff} = - \ln \bar{{\mathcal T}}.
\eq
Here $\bar{\mathcal T}$ denotes the average transmission over a single contiguous segment.  We will use to $\VEV{\tau_{\rm eff}}$ to denote the effective optical depth when averaged over many independent segments.

We therefore need a prescription relating $\Gamma$ to the local forest transmission.  If the IGM were a uniform medium, the relation would be trivial:  in ionization equilibrium, the \ion{He}{2} fraction obeys $x_{\rm HeII} \propto \Gamma^{-1}$, so $\tau \propto \Gamma^{-1}$ as well.  However, in the real IGM a forest of discrete clouds provides most of the absorption.  In this case, 
\bq
\tau_{\rm eff} = {1 + z \over \lambda} \int {\partial^2 N \over \partial N_{\rm HeII} \partial z} W_\lambda(N_{\rm HeII}) dN_{\rm HeII},
\label{eq:teff-forest}
\eq
where $\partial^2 N / \partial N_{\rm HeII} \partial z$ is the number of absorbers per unit redshift and per unit column density and $W_\lambda = \int (1 - e^{-\tau} ) d \lambda$ is the equivalent width of each line.  Because the abundances of lines in the \ion{He}{2} forest are poorly constrained at present, we write $N_{\rm HeII} = \eta N_{\rm HI}$ and use the observed column density distribution of the \ion{H}{1} forest to compute the integral; in the optically thin regime (relevant for the vast majority of low-column density systems providing \lya absorption), $\eta$ is independent of $N_{\rm HeII}$ \citep{haardt96, fardal98, faucher09}.  For such systems, a good approximation to the column density distribution is $\propto N_{\rm HI}^{-\beta}$, where $\beta \approx 1.5$ (\citealt{fardal98}, and references therein).  In that case, the integral in equation~(\ref{eq:teff-forest}) can be rewritten as \citep{fardal93},
\bq
\tau_{\rm eff} \propto \left( {\VEV{\Gamma} \over \Gamma} \right)^{\beta-1},
\label{eq:teff-prop}
\eq
where $\VEV{\Gamma}$ is the mean ionization rate.  Note that $\beta=0$ can mimic a uniform medium.  In fact, existing observations of the \ion{He}{2} forest are not of sufficient signal-to-noise to rule out the presence of a uniform component in the overall absorption, although most of the absorption does clearly come from discrete systems that can be associated with corresponding \ion{H}{1} absorbers (e.g., \citealt{fechner06}).  \citet{fardal98} showed that a completely uniform IGM roughly doubles fluctuations in the optical depth.

Of course, $\Gamma$ \emph{does} vary spatially throughout the simulation.  We, therefore, compute the transmission (relative to the mean) separately in each 1 Mpc pixel of the line of sight.  We then set the proportionality constant in equation~(\ref{eq:teff-prop}) so that the mean transmission $\VEV{\tau_{\rm eff}}$ has a specified value when averaged over our entire sample of lines of sight (usually chosen to match observations or an input model).  Finally, we compute $\tau_{\rm eff}$ averaged over larger continuous distances (typically $\sim 30$--$100 \Mpc$) by averaging over all the pixels within each such region.

\subsection{Other Sources of Fluctuations}
\label{otherflucs}

In the previous subsection, we showed how $\Gamma$ induces fluctuations in the transmission.  However, the \ion{He}{2} \lya forest is sensitive not only to the local radiation field but also to density fluctuations (the cosmic web) and to the IGM temperature field (which affects the recombination rate and, hence, the \ion{He}{2} fraction within each absorber).

To account for density fluctuations in the \ion{He}{2} forest, \citet{fardal98} considered variations in the counts of \ion{H}{1} \lya forest lines together with the transformation $N_{\rm HeII} = \eta N_{\rm HI}$.  They found that the fractional standard deviation of the effective optical depth of the forest over a segment of length (in redshift space) $\Delta z$ is\footnote{Here we use the forest-dominated case of \citet{fardal98}, rather than continuous IGM absorption.  We also use their model A2 for the distribution of \ion{H}{1} absorbers, which remains a reasonable fit to the low-column density absorbers relevant for the \ion{He}{2} forest.}
\bq
{\Delta \tau_{\rm eff} \over \VEV{\tau_{\rm eff}}} \approx {0.03 \over \sqrt{\Delta z}} ( \xi_{\rm HeII} \eta_{\rm 50})^{-0.1} \left( {1 + z \over 4} \right)^{-1},
\label{eq:dtaudensity}
\eq
where $\eta = 50 \eta_{50}$ is the ratio of column densities for \ion{He}{2} and \ion{H}{1} and $\xi_{\rm HeII}$ parameterizes the line width of the \ion{He}{2} lines relative to \ion{H}{1}:  it is unity for turbulent broadening (or that provided by the Hubble flow) and one-half for thermal broadening.  Theoretical models predict that the former is most relevant for the forest, but the measurements may be better fit by the latter \citep{fechner07-therm}.  Observations now show that $\VEV{\eta} \approx 50$--80 \citep{heap00, shull04, zheng04, fechner06, fechner07} with wide fluctuations between individual absorbing systems (that themselves depend upon the ionizing background and so are already incorporated in our model; \citealt{furl09-hefluc}).  Fortunately, the extremely weak dependence on $\xi_{\rm HeII}$ and $\eta_{50}$ implies that their uncertainties will not significantly impact our calculations.

Temperature fluctuations are substantial at the end of \ion{He}{2} reionization \citep{gleser05, furl08-igmtemp, mcquinn09}, and they will also affect the forest.  However, we expect their effects to be significantly smaller even than those due to density fluctuations.  We have  $x_{\rm HeII} \propto \Delta^2 \alpha(T)/\Gamma \propto \Delta^2 T^{-0.7}/\Gamma$, where $\Delta$ is the density in units of the mean and $\alpha(T)$ is the recombination rate.  The variance in $T$ is significantly less than an order of magnitude (and hence, smaller than or comparable to that in $\Delta$ and $\Gamma$), even at the height of \ion{He}{2} reionization, and is further damped by the smaller exponent. Thus that factor is much less important than the others, and we ignore it.

Finally, our simple Monte Carlo model also ignores the additional fluctuations sourced by the intricacies of radiative transfer through the IGM (for example, shadowing and spectral filtering by dense regions), as well as fluctuations in the absorber population.  For the latter, \citet{zuo93} showed that Poisson fluctuations in the absorber counts do not affect the correlation function of the ionizing background, so they should not affect our simpler statistics either.  The details of radiative transfer are best addressed with more sophisticated numerical simulations \citep{maselli05, tittley07, paschos07}.  We do not expect them to change our conclusions substantially, because we average over large spatial regions and so are not subject to the large variations that radiative transfer can induce on small scales. 

\subsection{The Attenuation Length}
\label{r0}

With the quasar luminosity function now well-measured (especially at the bright end), the key input determining the overall amplitude of the fluctuations in our model is the attenuation length of ionizing photons.  At $z=3$, estimates in the literature range from $\sim 30$--$60 \Mpc$ at the ionization edge \citep{bolton06, furl08-helium, faucher08-ionbkgd}, increasing by a factor of $\sim 50\%$ when averaged over the entire frequency spectrum.  Most studies take $R_0 \propto (1+z)^{-3}$ in the post-reionization limit, which we also assume as a fiducial model; this produces a reasonable match to the $z \la 2.7$ \ion{He}{2} \lya forest data \citep{dixon09}.  Of course, the overall level of fluctuations increases as the attenuation length decreases \citep{zuo92a, fardal93, meiksin03}, although the effect is not dramatic because of the wide range of intrinsic quasar luminosities \citep{meiksin07, furl09-hefluc}.  We will consider two fiducial attenuation lengths below, although we will not account for the much larger attenuation lengths of high-energy photons.  These will smooth out the background and so only strengthen our argument that the ``null hypothesis" of a smoothly evolving attenuation length is inconsistent with the observed fluctuation amplitude at $z \sim 3$.

\section{A Toy Model for Reionization}
\label{reion}

So far, we have assumed that \ion{He}{2} is highly-ionized everywhere, so that the only opacity source for high-energy photons is the (uniform) attenuation represented by $R_0$.  During reionization, however, large swathes of the IGM contain primarily \ion{He}{2}, absorbing nearly all photons above the ionization edge.  Such regions will, of course, appear as complete absorption troughs in the spectrum, naturally increasing the expected fluctuation amplitude.  

Unfortunately, a detailed model of the morphology of \ion{He}{2} and \ion{He}{3} regions requires a full numerical simulation of the reionization process, as in \citet{mcquinn09}.  Rather than attempt such a calculation, we will instead use a simple toy model to estimate the impact of reionization on the \ion{He}{2} \lya forest variations; we will, therefore, forego precise quantitative constraints during this era and, instead, use this model to demonstrate that reionization can plausibly explain many aspects of the observations. 

Specifically, we first choose the fraction $x_{\rm dark}$ of pixels that lie outside of regions illuminated by quasars.  This quantity is nearly, but not quite, the \ion{He}{2} fraction $x_{\rm HeII}$:  $x_{\rm dark} > x_{\rm HeII}$ because some regions that were ionized by earlier generations of quasars will no longer have a source illuminating them.  Without such a nearby source, recombinations are sufficiently rapid to quickly render $\tau_{\rm eff} \gg 1$ \citep{furl08-fossil, furl09-hefluc}.

We then assume that these opaque regions are precisely those with the smallest $\Gamma$ in the post-reionization limit; in other words, they are the regions least illuminated by \emph{existing} quasars.  We identify the threshold $\Gamma_{\rm dark}$ for which the probability that a random pixel has $\Gamma < \Gamma_{\rm dark}$ as $x_{\rm dark}$.  We then set $\Gamma = 0$ for all pixels below this threshold. We emphasize that this is an overly simplified prescription, because it does not account for the possibility that a \ion{He}{2} region may be relatively close to an existing quasar.  It is accurate in the limit in which quasars are extremely long-lived.  

Despite this rather extreme simplification, the model does have the nice property that brightly-illuminated regions near existing quasars are preserved (unlike if, for example, we laid down \ion{He}{2} regions randomly), which is qualitatively consistent with the evolving distribution of $\Gamma$ throughout reionization \citep{furl09-hefluc}.  Moreover, given that previously ionized regions will quickly become opaque due to recombinations \citep{furl08-fossil}, this simple model does provide the correct qualitative distribution.  It also does not require us to prescribe the spatial sizes of ionized regions.

\begin{figure*}
\plottwo{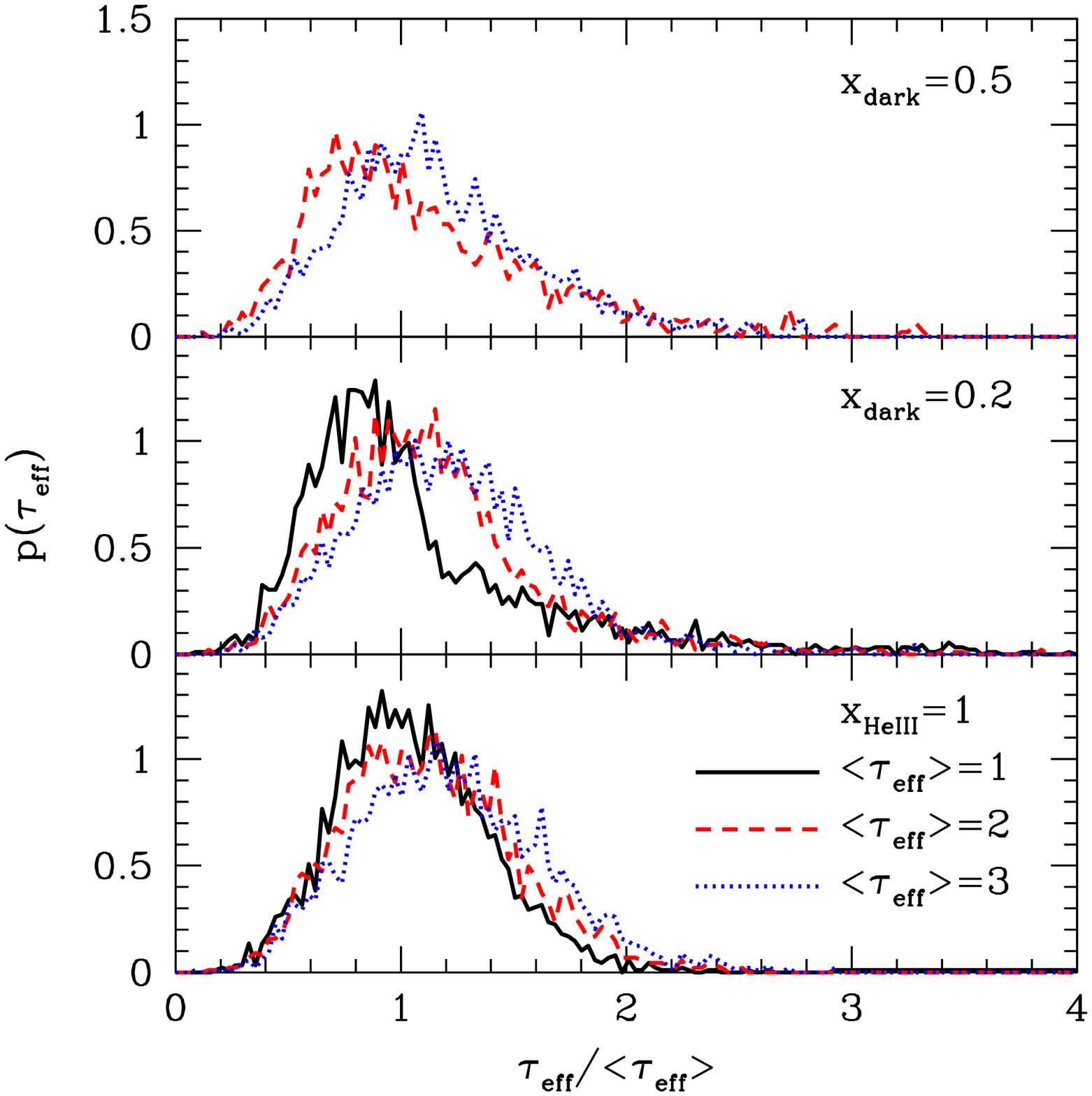}{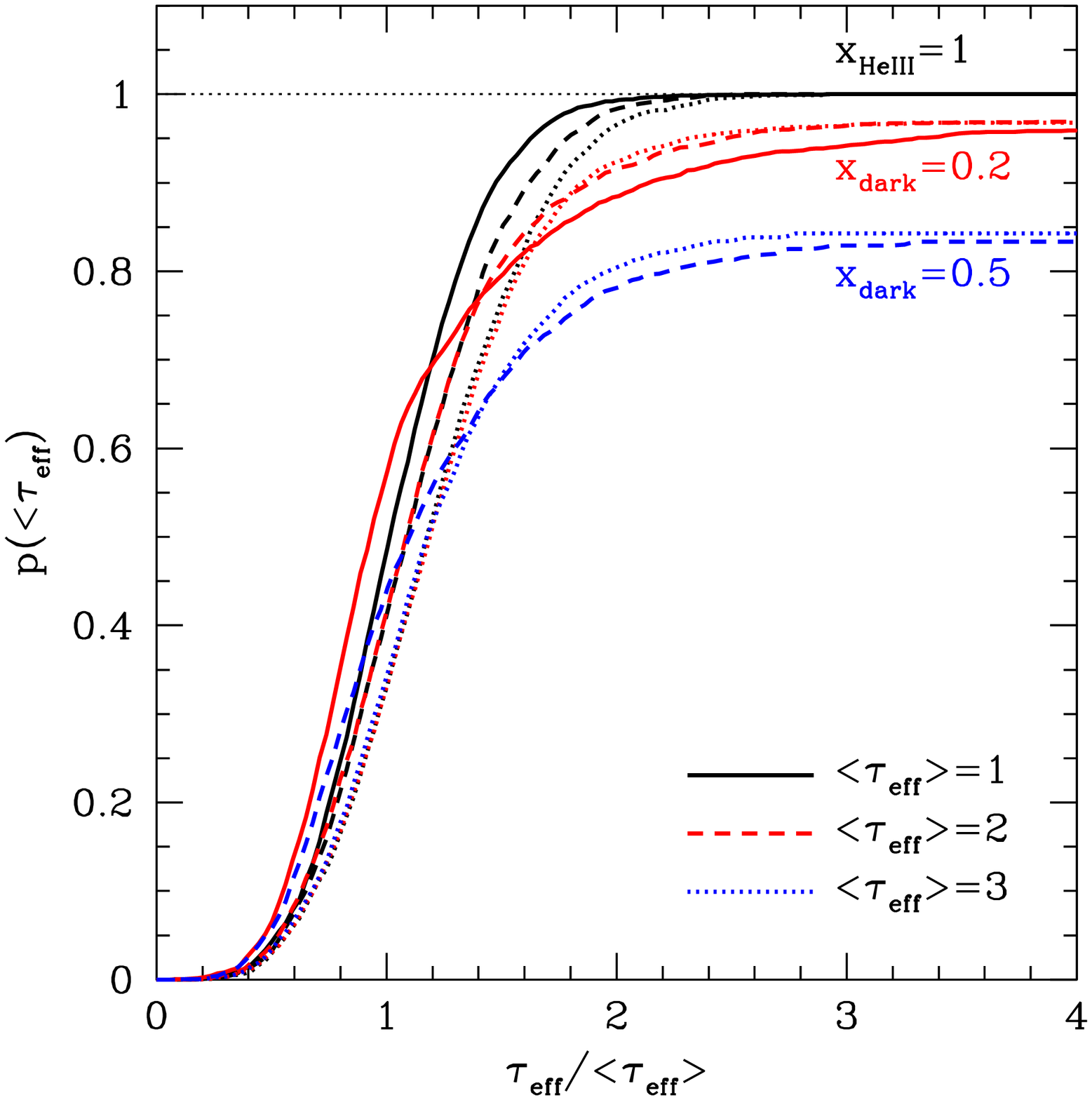}
\caption{\emph{Left:}  Probability distributions of $\tau_{\rm eff}$ averaged over 100 comoving Mpc segments in our simulations, scaled to an imposed mean value.  The three panels take different points during and after \ion{He}{2} reionization; the different curves normalize $\VEV{\tau_{\rm eff}}$ differently, as labeled in the bottom panel.  All are measured from our Monte Carlo model at $z=3$, assuming $R_0=30 \Mpc$.  \emph{Right:}  Same, but showing the cumulative probability distributions.
}
\label{fig:distbns}
\end{figure*}

Figure~\ref{fig:examples} shows some example lines of sight generated with this toy model if $x_{\rm dark}=0.2$.  The three original (post-reionization) lines of sight are shown in the bottom panel, to make clear the effects of our toy model.  One interesting aspect of the prescription is immediately obvious:  the dark regions span a wide range of scales, from $\sim 10 \Mpc$ to several hundred Mpc (from $r \sim 250$--$450 \Mpc$ along the dashed curve, for example).  Thus, at least in this toy model, both the early and late stages of reionization will enhance the observed fluctuations in $\tau_{\rm eff}$, even when we smooth the data on very large scales.

\section{Effective Optical Depth Distributions} 
\label{pdfs}

The left panels of Figure~\ref{fig:distbns} show the $\tau_{\rm eff}$ probability distributions for several simple cases.  Each takes segments of 100 Mpc (comoving), $z=3$, $R_0=30 \Mpc$, and uses our full Monte Carlo sample (with $3 \times 1$~Gpc$^3$ volumes, each with 100 lines of sight 1 Gpc long).  In each panel, we allow three different normalizations for $\VEV{\tau_{\rm eff}}$.  The right panel shows the cumulative distribution functions. We ignore density fluctuations in these models.

The post-reionization distributions (in the bottom left panel) contain substantial, but not overwhelming, variations, with a full width at half maximum comparable to the mean value.  In agreement with \citet{fardal98}, we find that the scaled distributions are relatively insensitive to $\VEV{\tau_{\rm eff}}$; our variance is also comparable to theirs, though our different choice for $R_0$ has some effect.  A larger mean opacity introduces more data points with moderately large $\tau_{\rm eff}$, because it is easier for a long IGM segment to lie entirely at low $\Gamma$.  However, the cumulative distributions make clear that none of these cases have regions with $\tau_{\rm eff}/ \VEV{\tau_{\rm eff}} \ga 2.5$ at a very high level of confidence (i.e., none appear in our 3,000 samples).

\begin{figure}
\plotone{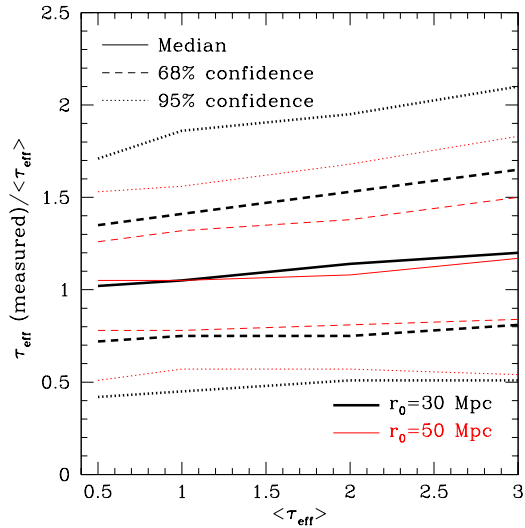}
\caption{Scaled distributions of measured $\tau_{\rm eff}$ values as a function of the true average opacity in the post-reionization limit.  The distributions average over 100 Mpc segments at $z=3$ and assume $R_0=30$ and 50 Mpc (thick and thin curves, respectively).  The solid curves show the medians of the distribution; the dashed and dotted curves show the limits within which 68\% and 95\% of the simulated measurements lie.}
\label{fig:errors}
\end{figure}

Figure~\ref{fig:errors} quantifies the post-reionization variations in a slightly simpler way.  Here, we show the medians of the distributions, as well as the limits within which 68\% and 95\% of the simulated segments lie.  First, note that the median is generally a good estimate of the average, although the skewness at larger $\VEV{\tau_{\rm eff}}$ does introduce a small bias.  The distribution remains relatively invariant toward small optical depths throughout the range in $\tau_{\rm eff}$.  Physically, strong transmission occurs near bright quasars, where the strong peaks in $\Gamma$ are independent of the average transmission.  On the other hand, the high-$\tau_{\rm eff}$ tail does broaden substantially at larger mean optical depths.  

Figure~\ref{fig:errors} also shows the impact of $R_0$ on the distributions; the thick and thin curves assume $R_0=30$ and 50 Mpc, respectively.  Overall, a larger attenuation length means more sources contribute to $\Gamma$ at each point in the IGM, so the stochastic fluctuations are smaller -- and, indeed, the distribution of $\tau_{\rm eff}$ narrows somewhat as $R_0$ increases.  However, the effect is not symmetric:  it is quite modest at $\tau_{\rm eff} < \VEV{\tau_{\rm eff}}$, because such segments are typically dominated by one or more luminous quasars, in whose immediate environs attenuation is unimportant.  On the other hand, the variations with $R_0$ are fairly large at $\tau_{\rm eff} > \VEV{\tau_{\rm eff}}$.  They are especially significant at the far end of this tail (the upper dotted curves), because it becomes very difficult to find long regions with extremely small values of $\Gamma$ as the volume sampled by each IGM point increases.

The other panels in Figure~\ref{fig:distbns} show the distributions from our toy model of \ion{He}{3} reionization (for $x_{\rm dark}=0.2$ and $0.5$ for the middle and top panels, respectively; in the latter case, we show only $\VEV{\tau_{\rm eff}}=2$ and 3, since a lower mean optical depth so early in reionization seems implausible).  Even in these cases, the bulk of the distribution does not change dramatically from the post-reionization limit.  However, a long tail toward high opacity clearly does develop; the cumulative distribution shows that $\sim 5\%$ and 15\% of the 100 Mpc skewers are completely dark for $x_{\rm dark}=0.2$ and $0.5$, respectively.\footnote{Actually, these regions will still have a weak ionizing background from X-ray photons, but this radiation field is not strong enough to render them transparent \citep{furl09-hefluc}.}  

Note that the observed fraction of dark segments is much smaller than $x_{\rm dark}$ itself -- because even a single bright quasar near the 100 Mpc skewer suffices to produce measurable transmission, if only over a short region -- but, even over long segments, a non-negligible fraction of completely dark segments remains at the end of reionization (see Fig.~\ref{fig:examples}).  Thus, at least in our simple models, reionization clearly can increase the measured variance, primarily by generating a tail toward extremely strong absorption, even after averaging the segments over large path lengths.  

In addition, when $\VEV{\tau_{\rm eff}}$ is relatively small, a more pronounced asymmetry does develop in the distribution, with the median falling somewhat farther below the average value (see, e.g., the solid curve in the middle panel on the left).  These cases have an exceptionally high contrast between the dark regions and those around luminous quasars (which provide almost all of the transmission).

So far, we have ignored fluctuations from the density structure (manifested through the line structure of the \lya forest); as shown by \citet{fardal98}, they are typically $\sim 0.1 \VEV{\tau_{\rm eff}}$ for segments of length 100 Mpc (or $\Delta z \approx 0.1$; see eq.~\ref{eq:dtaudensity}).  This is indeed a few times smaller than those from the radiation field, unless $R_0$ is many times larger than our fiducial values.

\section{Comparison to Data}
\label{data}

\begin{figure}
\plotone{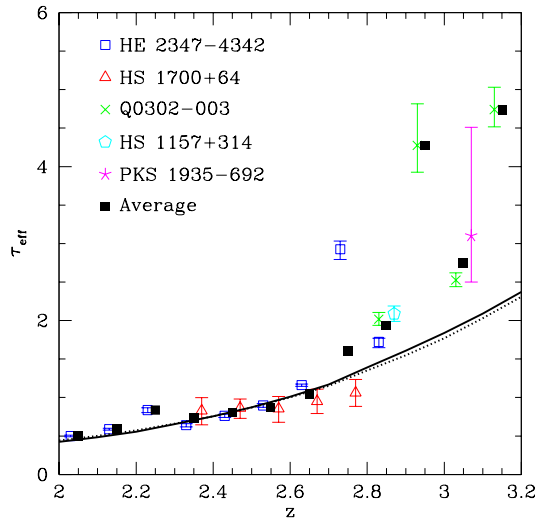}
\caption{Compiled \ion{He}{2} \lya forest effective optical depth measurements from the literature (see \citealt{dixon09} for references). The two curves are predictions of the optical depth assuming a smoothly evolving mean free path and the \citet{hopkins07} quasar luminosity function, normalized to the observed $\VEV{\tau_{\rm eff}}$ at $z=2.45$. The solid and dotted curves assume $R_0=30$ or $60 \Mpc$ at $z=3$, respectively. }
\label{fig:data}
\end{figure}

Although (to date) only five lines of sight have detailed \ion{He}{2} \lya forest spectra, the data still contain substantial fluctuations to which we can compare our model.  Figure~\ref{fig:data} shows all of the (averaged) $\tau_{\rm eff}$ measurements at $z < 3.2$ from the literature, as compiled by \citet{dixon09}. The data come from HE 2347-4342 \citep{kriss01, shull04, zheng04}, HS 1700+64 \citep{fechner06}, Q0302-003 \citep{hogan97, heap00, worseck06}, HS 1157+314 \citep{reimers05}, and PKS 1935-692 \citep{anderson99}.\footnote{We do not include the quasar discovered by \citet{zheng04-sdss}, because it provides only a lower limit to $\tau_{\rm eff}$ at $z \sim 3.5$.}  We have binned the measurements over $\Delta z=0.1$; within each bin, we show one data point for each line of sight, as well as the average over all the lines of sight (solid squares).  

\begin{deluxetable*}{ccccccc}
\tabletypesize{\scriptsize}
\tablecolumns{7}
\tablewidth{0pc}
\tablecaption{Probability Estimates for $R_0(z=3)=30 \Mpc$ \label{tab:conf-30}}
\tablehead{
\colhead{$\VEV{z}$} & \colhead{$\Delta r$ (Mpc)} & \colhead{$\tau_{\rm eff}$ (obs)} & \colhead{$\VEV{\tau_{\rm eff}}$ (model)} &
\colhead{P($x_{\rm dark}=0$)} & \colhead{P($x_{\rm dark}=0.2$)} & \colhead{P($x_{\rm dark}=0.5$)} 
}
\startdata
2.05 & 146 & $0.499^{+0.011}_{-0.012}$\tablenotemark{a} & 0.455 & 30\% & --\tablenotemark{b} & --\tablenotemark{b} \\
2.15 & 140 & $0.589^{+0.013}_{-0.013}$\tablenotemark{a} & 0.520 & 26\% & --\tablenotemark{b} & --\tablenotemark{b} \\
2.25 & 134 & $0.837^{+0.031}_{-0.034}$\tablenotemark{a} & 0.605 & 3\% & --\tablenotemark{b} & --\tablenotemark{b} \\
\hline
2.76 & 35 & $3.75^{+0.79}_{-0.43}$\tablenotemark{c,d} & 1.28 & $<0.01$\% & 10\% & 35\% \\
2.83 & 20 & $3.18^{0.58}_{-0.37}$\tablenotemark{c} & 1.50 & 1\% & 16\% & 33\% \\
\hline
2.93 & 61 & $4.37^{+1.13}_{-0.51}$\tablenotemark{e} & 1.72 & $<0.03$\% & 7\% & 25\% \\
3.07 & 58 & $2.53^{+0.09}_{-0.09}$\tablenotemark{e} & 1.90 & 34\% & 33\% & 45\% \\
3.15 & 94 & $4.74^{+0.46}_{-0.31}$\tablenotemark{e} & 2.23 & 2\% & 6\% & 17\% 
\enddata
\tablenotetext{a}{From HE 2347-4342 \citep{zheng04}.}
\tablenotetext{b}{We assume that helium reionization is complete by $z=2.3$ so do not report these values.}
\tablenotetext{c}{From HE 2347-4342 \citep{shull04}.}
\tablenotetext{d}{Note that \citet{kriss01} report a lower limit in this region of $\tau_{\rm eff} > 2.75$ with the same data.}
\tablenotetext{e}{From Q0302--003.  Measurements are from \citet{worseck06}, using data from \citet{heap00}.}
\end{deluxetable*}

The figure also shows two predictions for the evolution of $\tau_{\rm eff}$ assuming that helium reionization occurred at $z > 3.2$ (see \citealt{dixon09} for details).  Both curves take as inputs the quasar luminosity function from \citet{hopkins07} and a prescription for the attenuation length; we assume $R_0 \propto (1+z)^{-3}$ with $R_0=30$ and $60 \Mpc$ at $z=3$ for the solid and dotted curves, respectively.  These two values span the range of estimates in the literature for photons at the ionization edge (e.g., \citealt{bolton06, furl08-helium, faucher08-ionbkgd}).

With these input parameters, \citet{dixon09} combined the fluctuating radiation field models of \citet{furl09-hefluc} and the density field fit from \citet{miralda00} to generate $\VEV{\tau_{\rm eff}}$ via the fluctuating Gunn-Peterson approximation.  Because this approximation does not fully account for the line structure of the \lya forest, we have normalized both curves to the data point at $z=2.45$, which we regard as relatively robust.  This method allows us to compare the predicted optical depth \emph{evolution} to the observations, irrespective of any concerns about the normalization of the emissivity, the recombination rate, or the mean free path.  Because of this normalization, the two models yield very similar overall evolution for $\VEV{\tau_{\rm eff}}$, although of course the model with the larger attenuation length implies smaller overall fluctuations in the radiation field (and hence smaller variations in $\tau_{\rm eff}$).

There are several regimes that contain interesting discrepancies with our smooth model, as described in \citet{dixon09}, and we will examine each in detail below.  First, at $z \la 2.25$, the data from HE 2347-4342 imply that $\Gamma$ is $\sim 2$ times smaller than our simple extrapolation from higher redshifts would imply.  Second, at $z \sim 2.7$--2.9, the line of sight to HE 2347-4342 shows substantial fluctuations between large and small transmissions on $\ga 10 \Mpc$ scales.  Finally, at $z>2.9$, the line of sight to Q0302-003 shows a substantial increase in $\tau_{\rm eff}$ over the smooth models.  Thus we will focus on these two lines of sight (amongst the best observed) as particular instances of interesting fluctuations.  Our primary goal is to determine whether the data can be explained by including cosmic variance around the smoothly evolving model shown here.  If not, they imply substantial changes in the input parameters -- most likely \ion{He}{2} reionization.

Throughout the following, we will ignore variations in the underlying density field; equation~(\ref{eq:dtaudensity}) shows that such deviations are modest and do not substantially affect our conclusions. 

\subsection{Fluctuations at $z \la 2.3$}
\label{lowz}

We first wish to assess the significance of deviations from the ``smooth" $\Gamma$ evolution models at low redshift.  Near $z \sim 2$, \citet{dixon09} found that $R_0 \propto (1+z)^{-3}$ overpredicted the apparent ionizing background -- although only one line of sight has provided measurements in this regime (HE 2347-4342).  The deviations are much larger than the statistical errors from the observations; although in absolute terms, they only require a factor of two decrease in $\Gamma$.

We list the measured optical depths from HE 2347-4342 \citep{zheng04} and the predictions from our two models at $z<2.3$ in Tables~\ref{tab:conf-30} and~\ref{tab:conf-60}.  The first four columns in each table give the central redshift of the skewers, the comoving width of the skewers, the observed mean transmission (and $1\sigma$ statistical errors), and the predicted effective optical depth, respectively.  

Before evaluating whether these points are truly discrepant with the models, we do note two potential systematic problems with the data.  First, this low-$z$ regime corresponds to small wavelengths, so the continuum correction is most uncertain here.  Second, this regime has been corrected for Ly$\beta$ and higher-order line contamination using the corresponding \lya forest, but that correction may not be perfect given the noisy \ion{He}{2} data. Neither of these factors are included in the error estimates, so they already decrease the significance of the apparent discrepancy.

To estimate the significance of these deviations, we use our Monte Carlo model to generate 100 lines of sight in each of three different realizations of the quasar distribution.  Each box is ten times larger than the $\Delta z= 0.1$ segments used in the data binning.  We thus have 3000 total segments for each comparison.  We report the fraction of these model segments that have $\tau_{\rm eff}$ greater than or equal to the observed values in the fifth columns of Tables~\ref{tab:conf-30} and \ref{tab:conf-60}.  

\begin{deluxetable*}{ccccccc}
\tabletypesize{\scriptsize}
\tablecolumns{7}
\tablewidth{0pc}
\tablecaption{Probability Estimates for $R_0(z=3)=60 \Mpc$ \label{tab:conf-60}}
\tablehead{
\colhead{$\VEV{z}$} & \colhead{$\Delta r$ (Mpc)} & \colhead{$\tau_{\rm eff}$ (obs)} & \colhead{$\VEV{\tau_{\rm eff}}$ (model)} &
\colhead{P($x_{\rm dark}=0$)} & \colhead{P($x_{\rm dark}=0.2$)} & \colhead{P($x_{\rm dark}=0.5$)} 
}
\startdata
2.05 & 146 & $0.499^{+0.011}_{-0.012}$\tablenotemark{a} & 0.473 & 30\% & --\tablenotemark{b} & --\tablenotemark{b} \\
2.15 & 140 & $0.589^{+0.013}_{-0.013}$\tablenotemark{a} & 0.538 & 26\% & --\tablenotemark{b} & --\tablenotemark{b} \\
2.25 & 134 & $0.837^{+0.031}_{-0.034}$\tablenotemark{a} & 0.617 & 0.03\% & --\tablenotemark{b} & --\tablenotemark{b} \\
\hline
2.76 & 35 & $3.75^{+0.79}_{-0.43}$\tablenotemark{c,d} & 1.25 & $<0.01$\% & 10\% & 36\% \\
2.83 & 20 & $3.18^{0.58}_{-0.37}$\tablenotemark{c} & 1.45 & $<0.01$\% & 20\% & 35\% \\
\hline
2.93 & 61 & $4.37^{+1.13}_{-0.51}$\tablenotemark{e} & 1.66 & $<0.03$\% & 9\% & 28\% \\
3.07 & 58 & $2.53^{+0.09}_{-0.09}$\tablenotemark{e} & 1.96 & 18\% & 22\% & 45\% \\
3.15 & 94 & $4.74^{+0.46}_{-0.31}$\tablenotemark{e} & 2.17 & $<0.03$\% & 6\% & 20\% 
\enddata
\tablenotetext{a}{From HE 2347-4342.}
\tablenotetext{b}{We assume that helium reionization is complete by $z=2.3$ so do not report these values.}
\tablenotetext{c}{From HE 2347-4342 \citep{shull04}.}
\tablenotetext{d}{Note that \citet{kriss01} report a lower limit in this region of $\tau_{\rm eff} > 2.75$ with the same data.}
\tablenotetext{e}{From Q0302--003.  Measurements are from \citet{worseck06}, using data from \citet{heap00}.  }
\end{deluxetable*}

Clearly, the relatively small deviations at $z<2.2$ are not at all surprising in either model. However, the deviation at $z=2.25$ is significant, occurring in at most a few percent of the lines of sight. Such a discrepancy could more easily be accommodated if we allow fully dark regions (i.e., demand that \ion{He}{2} reionization is still ongoing), but we do not regard that as anywhere near likely.  

One obvious solution is that the model $\VEV{\tau_{\rm eff}}(z)$ falls off somewhat too steeply with redshift compared to the data.\footnote{As described in detail by \citet{dixon09}, another possibility is an evolving normalization factor in the fluctuating Gunn-Peterson approximation used there.}   Only a modest change would be necessary; increasing the model prediction by $\Delta \VEV{\tau_{\rm eff}}=0.05$ increases the expected probability to $\sim 15\%$ in the $R_0=30 \Mpc$ model.  Indeed, \citet{fechner06} find that the HS 1700+64 data are consistent with a smoothly evolving power law somewhat flatter than our curves, which would be completely consistent with the lower-redshift points (but make the disagreement at higher redshifts even more apparent).

Alternatively, density fluctuations may help reconcile the data.  Equation~(\ref{eq:dtaudensity}) suggests that this mechanism provides $\Delta \tau_{\rm eff} \sim 0.05$.  This kind of uncertainty can ultimately be calibrated by comparing to the \ion{H}{1} \lya forest.  However, in the case of the $z \sim 2.25$ region toward HE2347-4342, a visual inspection shows if anything \emph{more} \ion{H}{1} transmission than average \citep{zheng04}.

One final worry is that all three data points at $z < 2.3$ show a systematic increase in $\tau_{\rm eff}$ relative to the simple model, whereas one might expect fluctuations on both sides of the mean.  However, Figure~\ref{fig:examples} shows that it is not particularly uncommon to find regions several hundred Mpc long over which $\Gamma$ remains small.  To examine the likelihood of this, we consider estimates of the transmission over segments of length 420 Mpc at $z=2.15$.  We find that 7\% (2\%) of the model regions have $\tau_{\rm eff}$ above the observed value if $R_0=30 \ (60) \Mpc$. These are somewhat more likely than the $z=2.25$ segment on its own, so such large scale deviations are not particularly surprising.

\subsection{Fluctuations at $z \sim 2.8$}
\label{midz}

The second interesting region is at $z \sim 2.8$, where the HE 2347-4342 line of sight shows substantial fluctuations (both on its own and relative to the other lines of sight). We first focus on the segment at $z=2.76$, which spans $\sim 35 \Mpc$ and is nearly completely opaque (as measured most recently by \citealt{shull04}).  Here, we again use three independent boxes, 100 lines of sight in each box, and 30 short segments along each line of sight (for a total of 9,000 samples).  Tables~\ref{tab:conf-30} and \ref{tab:conf-60} show that neither of our post-reionization models can accommodate this point at a confidence level of $<0.01\%$.  Even if we use the reported $3\sigma$ lower limit $\tau_{\rm eff} > 2.78$, the probability is still less than 0.01\% in either model.  

On the other hand, if we allow for dark regions from incomplete helium reionization, the model probability becomes reasonably large: $\ga 10\%$ (35\%) for $x_{\rm dark}=0.2$ (0.5).  Thus, discounting any possible systematic errors in the data, this single point is \emph{extremely} suggestive that helium reionization has not yet completed.

Physically, this conclusion is so powerful, because the post-reionization models have a sharply defined minimum in $\Gamma$ throughout the Universe, generated by the accumulated background of quasars at large impact parameters (within two or three attenuation lengths).  This level is $\ga 0.3 \bar{\Gamma}$, so it is very difficult to accommodate large optical depths, even over relatively modest path lengths.  (On the other hand, excursions to small optical depths are easy to accommodate by passing near a bright quasar.)  

One possible explanation is a systematic error in the measurement. The relevant portion of the \emph{FUSE} spectrum is relatively close to an airglow emission line at 1136 \AA \ (corresponding to $z=2.73$), but overall this region is no more or less difficult to analyze than others in the spectrum (J. Tumlinson, private communication).  \citet{kriss01} examined the same data set and placed a $1\sigma$ lower limit of $\tau_{\rm eff} \ga 2.6$,\footnote{Note that this is averaged over a slightly different wavelength interval.} perhaps indicating the level of systematic uncertainties on the background subtraction, etc.  Nevertheless, the probability of finding such a segment in our post-reionization models is still $<0.1\%$.

Another possibility is that $\VEV{\tau_{\rm eff}}$ evolves more steeply at high redshifts, so that the mean value is actually closer to the measured value.  Figure~\ref{fig:data} shows that the measured average $\tau_{\rm eff}$ is also somewhat higher than our model.  However, if we assume that $\VEV{\tau_{\rm eff}}=1.75$ at $z=2.76$, somewhat above the measured average, we find that the probabilities for $\tau_{\rm eff}>3.75$ are $< 0.01\%$, although the probabilities for $\tau_{\rm eff}>2.75$ are more reasonable ($\sim 10\%$ and $1\%$ for the two mean free paths).  Thus a combination of the \citet{kriss01} analysis and an underestimate in the mean from our model would provide a plausible scenario.

However, as argued in \citet{dixon09}, it would be difficult to explain such a sudden upturn \emph{without} appealing to \ion{He}{2}  reionization.  The only possible causes are (1) rapid evolution in the ionizing emissivity or (2) rapid evolution in the attenuation length.  The first is ruled out by observations of the quasar luminosity function (e.g., \citealt{hopkins07}).  The second would  probably be associated with the continued clearing of IGM \ion{He}{2} during reionization, unless the modest increase in $\Gamma$ associated with reionization filters through the moderately dense \ion{He}{2} absorbers over a long time scale.

Of course, effects we have ignored in our model can also moderate the discrepancy somewhat.  For example, including density fluctuations (which, in our model, manifest themselves through the  number of \lya forest absorbers in a segment) increases the variations slightly, though not sufficiently to reconcile this measurement with the model.  More promising is radiative transfer:  one can imagine that shadowing, where a dense system blocks the light from a nearby source, could produce a regions with small $\Gamma$ and hence little transmission.  Still, because the lower limit on $\Gamma$ is provided by the collective action of many distant quasars and because the segment is relatively long ($\sim 35 \Mpc$), it seems difficult to arrange strong shadowing over the entire interval.

The same line of sight also contains strong fluctuations at $z=2.8$--$2.88$ toward both large and small transmission \citep{kriss01, shull04, zheng04}, with typical scales $\ga 10 \Mpc$.\footnote{These features are not shown in Figure~\ref{fig:data} because they are small-scale, and the \emph{average} opacity over the entire $z = 2.8$--2.88 segment toward HE2347-4342 is consistent with just a modest upturn from lower redshifts and is also consistent with other lines of sight at the same redshifts.  Nevertheless, the fluctuations are quite dramatic; see e.g. Figure~3 of \citet{dixon09}.} These features suggest similar conclusions, albeit at somewhat lower confidence.  In the table, we focus on a single measurement from \citet{shull04} at $z \approx 2.83$ that has a relatively large path length; the other strong fluctuations (toward both large and small transmission) are more easily accommodated because they have shorter path lengths.  

\subsection{Fluctuations at $z > 2.9$}
\label{highz}

We next consider the $z > 2.9$ range probed by Q0302-003.  In \citet{dixon09}, we found that these data imply a sharp break in $\VEV{\tau_{\rm eff}}$ at $z \la 2.9$, with substantially larger values at higher redshift.  For reference, we provide the binned $\tau_{\rm eff}$ measurements from \citet{worseck06}, who re-analyzed the \citet{heap00} observations, in Tables~\ref{tab:conf-30} and \ref{tab:conf-60}.  At $z=3$ (corresponding to 1216 \AA), geocoronal \ion{H}{1} \lya contaminates the spectrum; we have excised the range $z=2.96$--$3.04$ to conservatively account for this.  At $z=2.8$--2.9, the opacity along this line of sight is consistent with other measurements.  Above this range, the absorption increases sharply, then decreases (thanks largely to a nearby quasar discovered after the spectrum was taken; \citealt{jakobsen03}), and resumes increasing.

The last sections of Tables~\ref{tab:conf-30} and \ref{tab:conf-60} show the likelihood of finding the measured values in the context of our smoothly evolving models.\footnote{We do not list the $z=2.8$--2.9 region here because it is completely consistent with the smooth models and with the other lines of sight.}  Clearly, the point at $z=2.93$ is strongly discrepant with the post-reionization models, but it can be relatively easily accommodated if $x_{\rm dark} \ga 0.2$.  The same is true for the $z=3.15$ point.  On the other hand, at $z=3.07$, the substantial transmission spike from the nearby quasar makes the data entirely consistent with the smoothly evolving models. 

Of course, one can easily accommodate the measurements if our model underestimates $\VEV{\tau_{\rm eff}}$ at $z > 2.9$ by a significant factor, but as we have argued above, this itself would seem to require ongoing \ion{He}{2} reionization. In the end, reconciling the high- and low-$z$ behavior into smooth evolution appears difficult, requiring at least one model parameter to vary significantly more rapidly than typically assumed.

These conclusions are perhaps not surprising, given the other evidence indicating that \ion{He}{2} reionization occurs at $z \ga 3$.  However, the fact that two high-$\tau_{\rm eff}$ regions appear (out of four possible), when the probability of each is $\la 10\%$ at $x_{\rm dark} \sim 0.8$, perhaps also indicates that even $z=3$ is deeper in the reionization era than previously supposed.

\section{Discussion}
\label{disc}

We have computed the fluctuations expected in measurements of the average transmission of the \ion{He}{2} \lya forest due to large-scale variations in the radiation field. We used a simple Monte Carlo model of the quasar distribution to simulate the radiation field.  We then measured $\tau_{\rm eff}$ averaged over long ($\sim 30$--100 Mpc) segments along random lines of sight and computed its probability distribution.  For a fully-ionized Universe, the radiation field typically produces a full-width at half-maximum for the $\tau_{\rm eff}$ distribution of order its mean value \citep{fardal98}.  

We also constructed a toy model of reionization, where we demanded that a fraction $x_{\rm dark}$ of pixels have zero transmission.  We chose these dark pixels by identifying the regions that would have the smallest transmission in a post-reionization Universe.  Such a model is reasonable, because only active quasars can illuminate a region strongly enough to render it transparent, even if fossil, mostly-ionized regions fill much of the IGM \citep{furl08-fossil}. This method produces long dark stretches (with sizes $\ga 100 \Mpc$) in the empty voids between quasars, even late in reionization.  This procedure broadens the $\tau_{\rm eff}$ distribution by producing a large tail toward very high optical depths (and, to compensate, shifting the median transmission to smaller values).

We then compared our model to the observed fluctuations in the existing sample of \ion{He}{2} \lya forest lines of sight, as compiled by \citet{dixon09}.  In particular, we asked whether a post-reionization model with a smoothly evolving attenuation length (and hence no change in the ionization state of IGM helium) could account for the observed data.  

At lower redshifts ($z \sim 2.25$) such a model is reasonable, even though the data show somewhat stronger absorption than expected.  The observed transmission is unlikely according to our fiducial model, but modest changes to the normalization or slope of the underlying $\VEV{\Gamma}$ evolution can easily accommodate it.  Alternatively, the underlying IGM density fluctuations may also account for the discrepancy \citep{fardal98}.

This points to one important (though unsurprising) lesson from our analysis: even with the modest signal-to-noise already available in \ion{He}{2} \lya forest spectra, the statistical errors on $\tau_{\rm eff}$ are significantly smaller than the natural level of cosmic variance.  Thus, it is crucial to consider the latter when interpreting the existing data.

At higher redshifts, the data become much more difficult to explain.  Even though there are only five lines of sight with existing measurements, we found strong evidence that a smoothly-evolving post-reionization model is \emph{inconsistent} with the observations.  At $z \sim 2.7$--$2.9$, the measurements fluctuate strongly toward high optical depths on $\sim 20$--$35 \Mpc$ scales.  Such regions of deep absorption cannot fit into a post-reionization model (at $\ga 99\%$ confidence), because the large number of quasars far from the line of sight already provide enough radiation to keep the ionized fraction reasonably large.  Thus, the deviation cannot simply be explained as cosmic variance from a single line of sight, unless it is truly pathological.

To reconcile the observations and model, we need to increase the predicted $\VEV{\tau_{\rm eff}}$ significantly, use relatively small mean free paths, and take a conservative view of the data (i.e., accepting the more conservative \citealt{kriss01} lower limits).  However, in this case there \emph{still} must be a break in $\VEV{\tau_{\rm eff}}(z)$ at $z \sim 2.7$, which we have previously argued would likely itself signal \ion{He}{2} reionization \citep{dixon09}.

At $z \ga 2.9$, the only well-observed line of sight (Q0302-003) has $\tau_{\rm eff} \ga 4$ in two separate $\sim 60$--$100 \Mpc$ intervals.  This opacity is much larger than our smoothly-evolving model predicts, and even when post-reionization radiation fluctuations and density fluctuations are included, it remains extremely improbable.  Thus, this line of sight on its own also argues strongly for either ongoing reionization or a rapid increase in $\VEV{\tau_{\rm eff}}$ at $z \sim 2.9$, likely also signaling the end of reionization.  

Our model, therefore, supports the conclusion of \citet{dixon09}, who examined the evolving average effective optical depth, that the data are best explained by models with reionization completing at $z \la 2.9$.  This is somewhat later than expected from measurements of the IGM temperature \citep{schaye00, ricotti00, lidz09-temp}. It is more consistent with hints that the shape of the ionizing background spectrum changes around that time \citep{songaila98, songaila05}, although that argument has remained controversial \citep{kim02, aguirre04}.  Late reionization is reasonably consistent with estimates of the total ionizing emissivity of high-$z$ quasars, which produce a couple of ionizing photons per helium atom by this point (e.g., \citealt{furl08-helium}).

We have focused on variations in the radiation background. \citet{fardal98} showed that fluctuations induced by large-scale IGM density variations are modest.  While they will slightly modify the values in Tables~\ref{tab:conf-30} and \ref{tab:conf-60}, they will not qualitatively modify our conclusions about the high-$\tau_{\rm eff}$ measurements.  They can, however, help to reduce the tension at $z \la 2.3$, where the absolute deviations are relatively small.

Our toy model of reionization can obviously be much improved with realistic simulations of reionization, such as those performed by \citet{mcquinn09} -- although the large volume required to accumulate enough samples presents a challenge.  The probabilities generated from our simple model only provide a plausibility argument that this event, where large coherent regions with weak radiation fields should be relatively common, can more easily accommodate the observed $\tau_{\rm eff}$ distributions.  Qualitatively, we expect similar behavior in more sophisticated models (see, for example, the visualizations in Figs. 4-7 of \citealt{mcquinn09}).  We therefore have not tried to estimate the evolution of the \ion{He}{3} fraction with redshift but instead to argue that the data suggest some sort of change in the ionized fraction at $z \la 2.9$.

Our post-reionization model does require some caveats.  First, it does not include the details of radiative transfer, such as shadowing and spectral filtering.  These effects can clearly cause substantial fluctuations across the absorber population \citep{maselli05,tittley07}.  However, by averaging over large segments of each spectrum ($\ga 30 \Mpc$ across), we should also average over these details.  For example, it is difficult to imagine an absorber large enough to shadow an entire 30 Mpc region from the many sources that illuminate it within distances of a few attenuation lengths.

Second, we have ignored the deterministic clustering of quasars \citep{shen07}. \citet{furl09-hefluc} showed that this will be relatively unimportant to the distribution of $\Gamma$, because the fractional variance in quasar emissivity within one attenuation volume is much larger than that produced by deterministic clustering (thanks to the large Poisson fluctuations and the wide luminosity range of quasars).  \citet{mcquinn09} also showed that the distribution of \ion{He}{2} during reionization is essentially driven by the Poisson fluctuations of the quasar population.  However, quasar clustering may be somewhat more important for our application, because clustering \emph{within} the attenuation volumes cannot really be neglected \citep{mesinger09} and because the spatial coherence is important for averaging over long skewers.  However, it should still not be a dominant effect, and we expect that our qualitative conclusions will hold.

We have also simplified the problem by taking a single attenuation length for all photons, whereas in reality it increases with frequency.  However, the pair of values we have used spans a factor of two in length, both yielding similar conclusions, so we do not expect this to make a large difference.  If anything, including higher-energy photons will \emph{strengthen} evidence for late reionization, because larger attenuation lengths decrease the fluctuations.

It may seem surprising that we have drawn such strong conclusions, given the very limited number of lines of sight so far available.  The most conservative conclusion one could draw is that the data are inconsistent with a smoothly evolving ionizing background over the entire range $z \sim 2$--$3.2$.  Either (1) there must be a sudden increase in the slope of $\VEV{\tau_{\rm eff}}(z)$ or (2) extra fluctuations from reionization are required.  These conclusions are robust because they rely on the difficulty of producing a region  of any substantial length with $\Gamma \ll \bar{\Gamma}$ in the post-reionization Universe.  Thus, so long as an interval is large enough to include a substantial number of absorbers (and so average over the density fluctuations), the detection of even a single region with $\tau_{\rm eff} \gg \VEV{\tau_{\rm eff}}$ provides strong evidence for reionization.

In any case, the next several years should see a substantial increase in the number of lines of sight, with dozens of promising targets recently detected \citep{zheng04-sdss, zheng08, syphers09, syphers09b}.  The installation of the Cosmic Origins Spectrograph and repair of the Space Telescope Imaging Spectrograph on the \emph{Hubble Space Telescope} (HST) have provided two powerful instruments for such exploration; intriguingly, these instruments are best suited for studying the IGM at $z \ga 3$, precisely where the fluctuations should be most interesting.  \citet{syphers09, syphers09b} have detected 22 prime candidates for \ion{He}{2} \lya forest spectra.  Nearly all lie in the extremely interesting redshift range $z \sim 3$--$4$, in the heart of the reionization era.  A visual inspection of the modest-resolution spectra available from their surveys (taken with the ACS grism on HST) shows many interesting features both within individual spectra and across the entire population.  

Finally, our model shows that even modest-resolution and relatively low signal-to-noise investigations (such as the ACS grism sample) can make substantial progress in understanding \ion{He}{2} reionization.  We have drawn strong conclusions from the absorption averaged over fairly large intervals ($\Delta z \sim 0.03$--$0.1$), which is (at least in principle) a resolution-independent measurement.  Thus, this test can be useful even for quasars that are too faint for the detailed \ion{He}{2} \lya forest measurements that have so far held the focus of the community.

\acknowledgments

We thank J.~M. Shull, J. Tumlinson, and G. Worseck for sharing their data in electronic form.  This research was partially supported by the NSF through grant AST-0829737 and by the David and Lucile Packard Foundation.  We thank Agner Fog for making a public version of the Mersenne Twister algorithm available.


\end{document}